\documentclass[11pt]{article}
\usepackage[latin1]{inputenc}
\usepackage[T1]{fontenc}
\usepackage{amsmath}
\usepackage{amsfonts}
\usepackage{amssymb}
\usepackage{color}
\usepackage{graphicx}
\usepackage{tikz}

  \newcommand{\be}{\begin{equation}}
\newcommand{\ee}{\end{equation}}
\newcommand{\ba}{\begin{eqnarray}}
\newcommand{\ea}{\end{eqnarray}}

\newcommand{\bea}{\begin{eqnarray}}
\newcommand{\eea}{\end{eqnarray}}

\newcommand{\fs}
{i\kern+.01em\hbox{\raise.20ex\hbox{$/$}\kern-.58em$s$}}
\newcommand{\dslash}{\!\!\not\!\partial}

\newcommand{\bs} {i\kern-.01em\hbox{\raise.25ex\hbox{$/$}\kern-.52em$b$}}
\newcommand{\qs}{/\kern-.52em s}

\newcommand{\dd}
{\kern.06em\hbox{\raise.25ex\hbox{$/$}\kern-.40em$\partial$}}

\begin{document}

\tikzstyle{bag} = [text width=2em, text centered]
\tikzstyle{bag1} = [text width=5em, text centered]
\tikzstyle{end} = []
\title{Dualities and bosonization  of massless fermions  in three dimensional space-time}
\author{ E.~F.~Moreno$^a$  and
F.~A.~Schaposnik$^b$\thanks{Also at CICBA.} \\ \vspace{0.2 cm} \\
{\normalsize \it $^q$Department of Physics, Northeastern University}\\ {\normalsize \it Boston, MA 02115,
USA.} \\
{\normalsize \it $^b$\it Departamento de F\'\i sica, Universidad Nacional de La Plata}\\ {\normalsize \it Instituto de F\'\i sica La Plata}\\ {\normalsize\it C.C. 67, 1900 La Plata, Argentina}}

\date{\today}

\maketitle
\begin{abstract}
We study the bosonization of massless fermions in three-dimensional space-time. Using the path-integral approach as well as the operator formalism, we investigate new duality relations between fermionic and bosonic theories. In particular, we show that a theory of massless fermions is dual to three different, but equivalent bosonic theories which are quadratic in the bosonic fields: a non-local Maxwell- Chern-Simons-type theory, a non-local self-dual-type vector theory, and a local free massless bosonic theory. The equivalence is proven at the level of current correlation functions and current algebra
analysis.

\end{abstract}
\section{Introduction}
Bosonization of fermionic models in $d=3$ space-time dimensions has been discussed in the past following different approaches. Based in an order-disorder duality,  Marino \cite{Marino} was able to completely express a massless $d=3$ fermion field in terms of a bosonic vector field. In this way, fermionic current correlation functions can be reproduced in the framework of the bosonic theory described by a quadratic Lagrangian which  contains a Chern-Simons  and a nonlocal Maxwell-like term.

Concerning massive fermions, bosonization rules were established in ref.\cite{FS} within the path-integral framework. The dual bosonic Lagrangian, calculated to leading order in $1/m$ (with $m$ the fermion mass) was shown to correspond to  a Chern-Simons Lagrangian coupled to an ordinary (local) Maxwell term. The same result was obtained calculating the  one-loop quadratic part of the effective bosonic action, a method which is valid for arbitrary mass $m$ \cite{BFO} and reproduces in the $m\to 0$ limit the massless result. The current algebra of such quadratic   bosonic dual theory was proven to exactly reproduce the fermionic current algebra  \cite{LNS} and the extension to the non-Abelian case was   developed in \cite{GMNS}. Both in the massless and in the massive case the Chern-Simons term,  introduced in the pioneering works of Deser, Jackiw and Templeton \cite{DJT} plays a central role in all these results.

There has been recently a renewed interest in $d=3$ bosonization after  the report of new results on   Chern-Simons  theories coupled to scalars and fermion fields, in the large $N$ limit \cite{G}- {\cite{Dorigoni}} and their relation with Vassiliev higher spin theories \cite{V}-\cite{V2}. In particular, Aharony et al \cite{A} found, exploting the AdS/CFT correspondence,   that  the theory of $N$ massless scalars coupled to a  level $k$  $U(N)$ Chern-Simons term is equivalent, for large $N$,  to the Legendre transform of a theory of $k$ massless fermions coupled to a level $N$ $U(k)$ Chern-Simons term. Moreover, it has been conjectured that the equivalence of the two theories could be valid also at finite $N$   \cite{A}.

%

It is the purpose of this note to discuss the  fermion/boson connection in $d=3$ for massless fields using the approach developed in \cite{FS}-\cite{GMNS}. In particular, we search for new duality relations that may  provide a link with the results of ref.~\cite{A}. Our work is organized as follows. We review in section 2 the derivation of bosonization rules for massless fermions in $d=3$ space-time dimensions using the path-integral approach \cite{FS}-\cite{LNS}. We then derive in section 3 new dualities between the massless fermion theory and different bosonic theories with actions related to the Maxwell-Chern-Simons theory \cite{DJT}, the self-dual theory \cite{DJ} and a theory of decoupled scalars with higher derivative kinetic term. In section 4 we show the equivalence between one of the bosonic duals of massless fermions in $2+1$ dimensions and a massless spin zero bosonic model model; thus establishing a fermion-boson duality which resembles that found in \cite{A} particularized to the $U(1)$ case. Finally, we discuss our results in section 5.

\section{Bosonization of massless fermions in $3$ space-time dimensions in terms of vector fields}
We work within the path-integral approach focusing in the abelian case where the bosonization rule for fermion currents takes the form, in $d=3$ dimensional Euclidean space-time \cite{Marino},\cite{FS}
\be
j_\mu(x) \equiv i \bar\psi(x) \gamma_{\mu} \psi(x) \to -\frac1{\sqrt{4\pi}} \epsilon_{\mu \nu \alpha}
 \partial_\nu A_\alpha(x) \; , \;\;\;\; \mu=1,2,3
 \label{rule0}
 \ee
The vector field $A_\mu$ is the  bosonic field dual to the original fermion field.
An analogous formula for general space-time dimension $d$ was obtained in \cite{LNS}
 \be
j_\mu(x) \to C_d  \,\epsilon_{\mu_1\mu_2 \ldots \mu_d}
\partial_{\mu_2}\chi_{\mu_3 \ldots \mu_d}(x)
\label{ruled}
\ee
with $C_d$ a normalization constant and $\chi_{\mu_3 \ldots \mu_d}(x)$ an antisymmetric $d - 2$ Kalb-Ramond field in $d$ dimensional space-time \cite{Betal}. Note that  dualization of antisymmetric Kalb-Ramond fields in terms of derivatively coupled scalars is possible for all $d$ \cite{Fra}-\cite{Laba}, this could  open a route to $d=3$ bosonization  in terms of scalar bosons as in \cite{A}. We shall see, however that in this case the scalar field is trivially decoupled in the path-intgeral generating functional of current correletion functions, a result already discussed in \cite{Betal}.

It should be stressed that although the   bosonization rules (\ref{rule0})-\eqref{ruled} are exact, the  bosonic Lagrangian dual to the original free fermionic theory is not in general known in closed form since, as we shall see, this would require to have an exact expression for the determinant of the Dirac operator in a gauge field background. As it is well-known, only in the $d=2$ case the determinant can be calculated exactly. Concerning the $d=3$ case, we recall that the parity-odd contribution to such determinant -a Chern-Simons term- is known in closed form (see for example \cite{GRS}) while the parity-even one can only be calculated using some approximation method, namely order by order in  powers of the bosonic fields, or using the $\zeta$-function method in the case of  massless  fermions, or in powers of  $\partial/m$ for the case of fermions with mass $m$. Now, it is precisely the Chern-Simons term  that induces, on the bosonic side, the current bosonization rule given in eq.\,\eqref{rule0} and  a fortiori the current algebra. Moreover, that is the reason why it reproduces, in the bosonic dual theory, the exact Schwinger term result for the equal-time current commutators of the original $d=3$ fermionic theory calculated in \cite{dVG}.

Let us  briefly describe the main steps leading to eq.\,\eqref{rule0}. The partition function for $d=3$ free massless fermions in the presence of an external source $s_\mu$ for the fermion current is
\be
Z_\text{fer} [s_\mu] = \int D\bar\psi D\psi \exp\left(\int d^3x
\bar\psi \left(i  \dslash +  \not\! s \right)
\psi\right) \label{tres}
\ee
Performing a local change (Jacobian free) of the fermionic variables, $\psi(x) \to g(x)\psi(x)$, $g(x)\in U(1)$ gives
\ba
Z_\text{fer} [s_\mu] &=& \int D\bar\psi D\psi \exp\left(\int d^3x
\bar\psi \left(i  \dslash +  \not\! s + i g^{-1} \not \! \partial g \right)
\psi\right) \nonumber\\
&=&  \int Db_\mu D\bar\psi D\psi \exp\left(\int d^3x
\bar\psi \left(i  \dslash +  \not\! s +  \not \! b \right)
\psi\right) \delta[f_\mu[b]]
\label{delta}
\ea
where  $f_\mu(b) = \epsilon_{\mu \alpha \beta} \partial_\alpha b_\beta$ and the $\delta$-function in the second line ensures that $b$ is a flat connection, so that integration on $b_\mu$ reproduces the first line.

Integrating out fermions and after the trivial shift $b_\mu + s_\mu \to b_\mu$
the partition function becomes
\be
Z_{fer}[s] = \int Db_\mu \det\left(i  \dslash +  \not\! b
\right)  \delta[f_\mu[b] - f_\mu[s]]
\label{deltas}
\ee
It is at this point that the bosonic field dual to the original fermion enters into play. Indeed, representing the delta function in \eqref{deltas} in the form
\be
\delta[f_\mu[b] - f_\mu[s]] = \int DA_\mu \exp\left(i\int d^3x  (f_\mu[b] - f_\mu[s])A_\mu
\right)
\ee
the fermionic partition function can be finally written as a bosonic partition function $Z_{bos}[s]$
\be
Z_{fer}[s] = Z_{bos}[s]
\ee
where
\be
Z_{bos}[s] \equiv \int DA_\mu \exp(-S_{bos}[A])  \exp\left(-i\int d^3x f_\mu[s] A_\mu)
\right) \label{ocho}
 \ee
with the bosonic action  given by
\be
\exp(-S_{bos}[A]) = \int Db_\mu \det (i \dslash + \not b)
\exp\left(i\int d^3x f_\mu[b] A_\mu\right ) \label{nueve}
\ee
We see that the  bosonic field $A_\mu$, dual to the original fermion field, arises as the Lagrange multiplier enforcing the flatness condition for the auxiliary field $b_\mu$ in eq.\eqref{deltas}

Eq.~(\ref{ocho}) shows that the external source $s_\mu$ has factored out from the auxiliary field $b_\mu$ integration so that already at this stage  {one  can} obtain the exact bosonization rule \eqref{rule0} by differentiating $Z_{bos}[s]$
\be
\langle j_\mu(x)\rangle =\left. \frac{1}{Z}\frac{\delta Z}{\delta s^\mu(x)}\right|_{s=0} =  \langle-\frac1{\sqrt{4\pi}} \epsilon_{\mu \nu \alpha}
\partial_\nu A_\alpha(x)\rangle
\label{rule00}
\ee
Of course one has still to determine the bosonic action $S_{bos}[A]$ so as to be able to effectively compute the vacuum expectation value in the  r.h.s. in \eqref{rule00}.

As already pointed out above,   the derivation we have presented can be easily extended to any number of space-time dimensions, in which case instead of the   vector field $A_\mu$  the field dual to the $d$-dimensional fermion will be {an} antisymmetric Kalb-Ramond field $\chi_{\mu_3 \ldots \mu_d}$  with the bosonization rule for the fermion current given in eq.\,\eqref{ruled}. Again, the dual bosonic field $\chi_{\mu_3 \ldots \mu_d}$ arises as a Langrange multiplier with its rank determined from the form of the flatness conditions of the auxiliary field-strength $f_{\mu\nu}[b]$ in different   space-time dimensions.

It is interesting to note that  in  the $d=2$ case  with an appropriate choice of constant $C_2$ the bosonization rule eq.~\eqref{ruled}
takes well-honored form \cite{Coleman}
\be
j_\mu(x) \to -\frac1{\sqrt\pi} \epsilon_{\mu\nu} \partial_\nu\phi(x)
\ee
The r.h.s. of this formula is the topological current in $d=2$ space-time dimensions constructed from a
bosonic scalar $\phi$. Analogously,  the r.h.s. in eq.~\eqref{rule0} is nothing but the topological current in $d=3$ dimensions for the vector field $A_\mu$.

As explained above, the parity-even part of the fermion determinant appearing in the definition of $S_{bos}[A]$, eq.\eqref{nueve}, cannot be computed in closed form for $d>2$. Making an expansion in powers of the field $b_\mu$  and computing up to second order terms one can find, either using the order-disorder operator approach proposed in ref.\,\cite{Marino} or the path-integral approach developed in ref.\,\cite{LNS},  
\be
S_{bos}[A] =   \int d^3x \left(\frac{1}{2\pi} F_{\mu \nu}
T^{-1} F_{\mu \nu} -  {\frac{i}\pi}  \epsilon_{\mu \nu \lambda}
A_\mu \partial_\nu A_\lambda \right) \equiv S_{TMCS}[A]
\label{m}
\ee
We again stress that the Chern-Simons parity violating term does not pick up higher order corrections. Concerning the Maxwell-like term, which does receive corrections, we denote with $T^{-1}$  the Green function of the square root of the (minus) Euclidean d'Alambertian (i.e., the $d=3$ Laplacian),
\be
T\equiv
(-\Box  )^{1\!/2} \; , \;\;\;\; (T*T^{-1})_{x,y} = \delta^{(3)}(x-y) \, .
\label{T operator}
\ee
Since we are working in Euclidean space one has
\be
\Box = \partial_0^2 + \nabla^2 \, , \;\;\;\; \nabla^2 = \partial_{x_1}^2 + \partial_{x_2}^2
\ee

From the identity
\be
- \Box \exp(ikx) = |k|^2 \exp(ikx)
\label{explained}
\ee
and  the natural definition of the square root of the Laplacian it can be established that
\be
T  \exp(ikx)  =  |k| \exp(ikx)
\label{valorabs}
\ee
and
 \be
T^{-1}(x,0) = \int \frac{d^3k}{(2\pi)^3} \frac1{|k|} \exp(ikx)
  \ee
A rigorous definition and properties of the square root of Laplacians  can be found in \cite{Caffa}-\cite{MS}.

Were fermions massive, one would have a Maxwell term  $(1/(|m|) F_{\mu\nu}^2$ instead of the  first term in \eqref{m}. In this respect, since in the massless fermionic case there is no dimensionful parameter, it is natural to find in  the second order term of the even-parity contribution to the fermion determinant a $T^{-1}$ factor leading to the correct dimensionless Maxwell-like term.

Note that while the fermionic action  is invariant under global $U(1)$ transformations, the bosonic action exhibits a  local $U(1)$ gauge  invariance, provided $A_\mu$ is endowed with appropriate boundary conditions. This invariance is a consequence of the natural choice of the gauge-invariant procedure adopted in the path-integral bosonization approach that we employed and also arises within the order-disorder operator algebra approach of ref.\,\cite{Marino}. As it will be discussed below, there is a regularization ambiguity parameter $\alpha $ that for the moment we have chosen as $\alpha=i$ in the previous formul\ae. It should be stressed that, in spite of being non-local, action $S_{TCSM}$ is perfectly manageable. Indeed, the choice of a retarded prescription shows that the non-locality preserves causality and a canonical quantization of the theory can be formulated \cite{Amaral}.

Returning to the bosonization rule \eqref{rule0}, one can easily see  that it reproduces the correct commutation  relations, both  for the fermionic and the bosonic dual theories. Indeed, the well-known current-current commutators for fermions giving a Schwinger term, which in $d=3$ takes the form
\cite{dVG}
\be
[j_0(\vec x,t),j_i(\vec x,t)] = -\frac1{8\pi} \lim_{\epsilon \to 0} \frac1\epsilon \partial_i \delta^{(2)}(\vec x - \vec y)
\ee
can be obtained from the bosonic  theory if, according to \eqref{rule0} one identifies
\begin{align}
j_0 \to \frac1{\sqrt{4\pi}} B \;, \quad j_i \to \frac1{\sqrt{4\pi}} \epsilon_{ij} E_j
\end{align}
and uses the bosonic action $S_{TMCS}$ to compute current commutators using the Bjorken-Johnson-Low approach. We have written $E_i = F_{i0}$ the electric field and $B = \epsilon_{ij}\partial_i A_j$ ($i,j=1,2$) the magnetic field. Interestingly enough, from this point of view, the origin of the Schwinger term in the free fermion theory can be traced back to the appearance of a Chern-Simons term in the dual bosonic action (see \cite{CDFKS} for a recent discussion on this issue).
\section{Dualities among bosonic actions}

We shall now see that the action $S_{TCSM}$ is dual to a self-dual bosonic action. This was proven for massive  fermions in \cite{FS}; here we extend the result to the massless case. To this end, let us start by considering the  Maxwell-like term in the bosonic action \eqref{m}
\be
\frac1{2\pi}\int d^3x\,  F_{\mu\nu}[A] T^{-1} F_{\mu\nu}[A] =  {\frac{1}{2}} \int d^3x \,J_\mu[A] T^{-1} J_\mu [A]
\label{14}
\ee
where
\be
J_\mu[A] = {\sqrt{\frac{2}{\pi}}}\varepsilon_{\mu\nu\alpha} \partial_\alpha A_\beta
\ee
Using the identity
\be
\exp\left(- \int d^3x
 J_{\mu}[A]
T^{-1} J_{\mu}[A] \right) = \int Da_\mu \exp \left(- \frac12a_\mu T a_\mu + a_\mu J_\mu[A]
\right)
\label{martes}
\ee
 $Z_{bos}[s]$ can be written as
\be
Z_{bos}[s] = \int DA_\mu Da_\mu \exp\left( {+} \frac12a_\mu T a_\mu {-} a_\mu J_\mu[A] + \frac{i}\pi  \epsilon_{\mu \nu \lambda}
A_\mu \partial_\nu A_\lambda +  {\frac{1}{\sqrt{4\pi}} \epsilon_{\mu \nu \lambda}
s_\mu \partial_\nu A_\lambda} \right)
\ee

If we now proceed to integrate over $A_\mu$ we get
\begin{align}
\int DA & \exp\left(- \int d^3x  \left(\frac{i}\pi \epsilon_{\mu\nu\alpha}A_\mu \partial_\nu A_\alpha -  { {{2i}}\epsilon_{\mu\nu\alpha}\left( a_\mu -\frac1{\sqrt{8}} s_\mu\right) \partial_\nu  A_\alpha}
\right)\right) = \\
 & {\exp \left(-  \frac{i}{2} \int d^3x  \epsilon_{\mu\nu\alpha}\left(a_\mu \partial_\nu a_\alpha -\frac{2}{\sqrt{8}}s_\mu \partial_\nu a_\alpha -\frac{1}{8} s_\mu \partial_\nu s_\alpha \right) \right)}
\end{align}
so that we end with a generating functional of the form
\ba
Z_{bos}[s] &=&  \int Da_\mu \exp\left(- \int d^3x\left(\frac12a_\mu T a_\mu +   \frac{i}2  \epsilon_{\mu \nu \lambda}
a_\mu \partial_\nu a_\lambda\right)\right) \nonumber\\
&& \times \exp\left(
 -\int d^3x \left(\frac{1}{{\sqrt2}}s_\mu \partial_\nu a_\alpha +\frac{1}{8} s_\mu \partial_\nu s_\alpha \right) \right)
 \equiv Z_{TSD}[a;s]
 \label{20}
\ea
In this way we have proven at the quantum level an exact duality identity for the    following bosonic actions
\be
S_{TMCS} = \frac{1}{2\pi}  \int d^3x
 \left( F_{\mu \nu}
T^{-1} F_{\mu \nu} -
 {2i}   \epsilon_{\mu \nu \lambda}
A_\mu \partial_\nu A_\lambda\right) \rightleftarrows
S_{TSD} = \int d^3x \left(\frac12a_\mu T a_\mu +   \frac{i}2  \epsilon_{\mu \nu \lambda}
a_\mu \partial_\nu a_\lambda \right)
\label{dualT}
\ee
to be compared with the quantum duality between selfdual ($SD$) and Maxwell-Chern-Simons ($MCS$) models established in \cite{DJ} within a canonical approach and in \cite{FS},\cite{S} within the path integral approach in the massive case,
\be
S_{MCS} = \frac{1}{2\pi} \int d^3x\left(\frac{1}{|m|}\, F_{\mu \nu}F_{\mu \nu} -
{2i}  \epsilon_{\mu \nu \lambda} A_\mu \partial_\nu A_\lambda\right) \rightleftarrows
S_{SD} = \int d^3x\left( \frac12 m a_\mu a_\mu +\frac{i}\pi  \epsilon_{\mu \nu \lambda}
a_\mu \partial_\nu a_\lambda\right)
\label{dual}
\ee

Let us end this section by describing a different dualization that is possible due to the fact that one can always  trade  antisymmetric Kalb-Ramond fields (in this case just a vector field) to scalars with a higher derivative Lagrangian\cite{Fra}-\cite{Laba}. Following \cite{Betal} we start from eq.(\ref{deltas}) and add a gauge fixing term for $b_\mu$,
\be
Z_\text{fer} [s_\mu] = \int D b_\mu \, \text{det}\left(i \dslash +  \not\! s + \not \! b \right)
\, \delta (f_\mu(b))\, e^{\frac{1}{2\zeta} \int d^3 x\, (\partial \cdot b)^2}
\ee
To enforce the constraint we introduce as before the vector  field $A_\mu$,
\begin{align}
Z_\text{fer} [s_\mu] = \int D A_\mu D b_\mu \, \text{det}\left(i \dslash +  \not\! s + \not \! b \right)
\, e^{i \int d^3 x \, A_\mu f_\mu(b)} \, e^{\frac{1}{2\zeta} \int d^3 x\, (\partial \cdot b)^2}
\end{align}
The first exponential factor  in the r.h.s. can be written as $\exp\left({i \int d^3 x \, b_\mu f_\mu(A)}\right)$ making evident that the field $A_\mu$ is also a gauge field. Shifting $b_\mu \to b_\mu - s_\mu$ we absorb the source in the determinant obtaining
\begin{align}
Z_\text{fer} [s_\mu] = \int D A_\mu D b_\mu \, \text{det}\left(i \dslash +  \not \! b \right)
\, e^{i \int d^3 x \, A_\mu (f_\mu(b)-f_\mu(s)} \, e^{\frac{1}{2\zeta} \int d^3 x\, (\partial \cdot (b-s))^2}
\end{align}
The next step is to integrate the field $b_\mu$.  To this end  we shall work to lowest order in the fields so that the fermion determinant becomes
\be
\text{det}\left(i \dslash +  \not \! b \right) = \text{exp}\left\{ - \int d^3x\, b_\mu \, \Pi_{\mu \nu} \, b_\nu\right\}
\ee
where $\Pi_{\mu\nu}$,  is the $d=3$  vacuum polarization,
\be
\Pi_{\mu\nu} = \frac{1}{\pi}\left(\partial_\nu T^{-1} \partial_\mu - \delta_{\mu \nu} \partial_\alpha T^{-1} \partial_\alpha \right) +\frac{1}{\pi} \epsilon_{\mu \nu \alpha}\partial_\alpha
\ee
As it should be, the vacuum polarization is transverse, $\partial_\mu \Pi_{\mu\nu} =0$.

Substituting the value of the fermion determinant in the partition function, we get
\begin{align}
Z_\text{fer} [s_\mu] = \int D A_\mu D b_\mu \, e^{-S[b,A,s]}
\end{align}
where
\begin{align}
S[b,A,s]= \int d^3x \, \left\{ b_\mu \, \hat \Pi_{\mu \nu} \, b_\nu + b_\mu \left( - i f_\mu(A) + \frac{1}{\zeta} \partial_\mu \partial \cdot s \right) + i s_\mu f_\mu(A) - \frac{1}{2\zeta} s_\mu \partial_\mu \partial_\nu s_\nu \right\}
\end{align}
and $\hat \Pi_{\mu \nu} = \Pi_{\mu \nu}  -\frac{1}{2\zeta} \partial_\mu \partial_\nu$.
Since this action is quadratic, the path-integral over $b_\mu$ can be performed in close form leading to
\begin{align}
Z_\text{fer} [s_\mu] = \int D A_\mu \, e^{-S_2[A,s]}
\end{align}
where
\begin{align}
S_2[A,s]= \int d^3x \, &\left\{ \frac{1}{4}\left(f_\mu(A) \, \hat \Pi_{\mu \nu}^{-1} \, f_\nu(A)\right)  + i\, f_\mu(A)\left(\frac{1}{2\zeta} \hat \Pi_{\mu \nu}^{-1} \partial_\nu (\partial \cdot s )+ s_\mu\right) - \right. \nonumber\\
& \left.  \frac{1}{4\zeta^2} \left(\partial_\mu (\partial \cdot s) \hat \Pi_{\mu \nu}^{-1} \partial_\nu (\partial \cdot s)\right)-\frac{1}{2\zeta}\left(s_\mu \partial_\mu \partial_\nu s_\nu\right)
\right\}
\end{align}
Now we trade the integration over $A_\mu$ by an integration over $f_\mu(A)$ using the identity
\be
\int DA \, F[f_\mu(A)] = \int Df\, F[f]\, \delta(\partial \cdot f)
\ee
Introducing  a scalar field $\varphi$ to enforce the $\delta$-function, we then get
\begin{align}
Z_\text{fer} [s_\mu] = \int D f_\mu D\varphi \, e^{-S_3[f,\varphi,s]}
\end{align}
with
\begin{align}
S_3[f,s]= \int d^3x \, &\left\{ \frac{1}{4}\left(f_\mu \, \hat \Pi_{\mu \nu}^{-1} \, f_\nu\right)  + i\, f_\mu\left(\frac{1}{2\zeta} \hat \Pi_{\mu \nu}^{-1} \partial_\nu (\partial \cdot s )+ s_\mu +\partial _\mu \varphi\right) - \right. \nonumber\\
& \left.  \frac{1}{4\zeta^2} \left(\partial_\mu (\partial \cdot s) \hat \Pi_{\mu \nu}^{-1} \partial_\nu (\partial \cdot s)\right)-\frac{1}{2\zeta}\left(s_\mu \partial_\mu \partial_\nu s_\nu\right)
\right\}
\end{align}
The integral over $f_\mu$ is gaussian and can be performed, leading to
\begin{align}
Z_\text{fer} [s_\mu] = \int D \phi \, e^{-S_\text{eff}[\varphi,s]}
\end{align}
with
\begin{align}
S_{eff} &= \int d^3x \left\{ \left(s_\mu + \frac{1}{2\zeta} \hat \Pi^{-1} _{\mu \alpha} \partial_\alpha \partial \cdot s + \partial_\mu \varphi\right) \hat \Pi_{\mu \nu} \left(s_\nu + \frac{1}{2\zeta} \hat \Pi^{-1} _{\nu \beta} \partial_\beta \partial \cdot s+\partial_\nu \varphi\right) + \right.\nonumber\\
&\left. \frac{1}{2\zeta}\, s_\mu \partial_\mu \partial_\nu s_\nu + \frac{1}{4\zeta^2}\, s_\mu \partial_\mu \partial_\alpha \hat \Pi^{-1}_{\alpha \beta} \partial_\beta \partial_\nu  s_\nu\right\}
\end{align}
Expanding this last expression, and using that $\Pi_{\mu \nu}$ is transverse, we finally have
\be
Z_{fer}[s_\mu] =\int D\varphi \text{exp}\left\{ - \int d^3 x \left( s_\mu \Pi_{\mu \nu} s_\nu - \frac{1}{2\zeta} \varphi \Box\Box
\varphi \right) \right\}
\ee
The source is decoupled from the field $\varphi$ which has a higher derivative propagator, a result which is identical to the one  found in \cite{Betal} in any number of dimensions including $d=3$. One should note that the $s_\mu$ term reproduces the correct result for the fermionic current algebra so that, as observed in \cite{Betal} its functional integral may   be completely ignored, contributing as it does just a field-independent overall constant.

\section{Bosonization in terms of a canonical free field $\phi$}

It has been proven in \cite{DJT}   that the field equations and the commutation relations  of the massive  quantum  Maxwell-Chern-Simons theory are solved by a canonical, free, massive spin 1 field (with the mass coinciding with the Chern-Simons action coefficient). Following  the approach developed in  \cite{DJT}, we will show the equivalence between the bosonic model with action $S_{TMCS}[A]$ defined in equation \eqref{m} and a massless free field model. Our results can be seen as  establishing a fermion-boson duality which agrees, in the $U(1)$ case with that discussed in \cite{A}.

Let us start from our the bosonized Lagrangian containing Maxwell-like and a Chern-Simons actions

\be
S_{TMCS}[A] =  \int d^3x \left(\frac{1}{2} F_{\mu \nu} T^{-1} F_{\mu \nu} -
 \alpha\,  \epsilon_{\mu \nu \lambda} A_\mu \partial_\nu A_\lambda \right)
\label{7}
\ee
Here  we have reintroduced the regularization parameter $\alpha$
previously fixed to a particular value. The Gauss law derived
from action \eqref{7} takes the form
\be \partial_i (T^{-1} E_i)=-2 \alpha B
\ee
where as before  $E_i=F_{i0}$ and $B=\epsilon_{ij} \partial_i A_j$.

Following \cite{DJT} we now propose to solve this constraint in terms of a canonical free field $\phi$,
\begin{align}
E_i &= \epsilon_{ij} \mathbf{A} \,\partial_j\Pi + \mathbf{B}\, \partial_i\, \phi \nonumber\\
B &= \mathbf{C} \, \phi
\label{gauss}
\end{align}
where $\mathbf{A}$, $\mathbf{B}$, and $\mathbf{C}$,  are three kernels to be determined.
Using this expression to write the Gauss law  we obtain
\[\partial_i (T^{-1}*E_i)=\nabla^2 T^{-1}*\mathbf{B}*\phi \]
so
\be
\nabla^2 T^{-1}*\mathbf{B}=-2\, \alpha\, \mathbf{C}
\label{box}
\ee

Given action \eqref{7}, the corresponding Hamiltonian reads
\[ H=\frac{1}{2} \int d^2x \left( E_i\,T^{-1}\,E_i + B\,T^{-1}\,B\right) \]

Writing ${\mathbf{B}=a\, T^{3/2} \hat T^{-1}}$ and choosing $\hat T=(-\nabla^2)^{-1/2}$ with  $a$ an arbitrary parameter one can see that the magnetic term in the Hamiltonian can be written as
\begin{equation}
\int d^2x B\,T^{-1}\,B  = - \frac{a^2}{4\alpha^2}\int d^2x \,\phi \nabla^2 \phi
\end{equation}
Note that $\hat T$ contains solely spacial derivative. To obtain this result we have used that $T$ and $\hat T$ commute (which can be proven following  \cite{Caffa}-\cite{MS}).

Concerning the electric contribution to the Hamiltonian,
\begin{align}
\int d^2x\, E_i\,T^{-1}\,E_i &= \int d^2x \, (\mathbf{B}\,\partial_i\phi)(T^{-1}\,\mathbf{B}\,\partial_i\phi) + 2\, \int d^2x \,(\epsilon_{ij} \mathbf{A}\,\partial_j\Pi)(T^{-1}\,\mathbf{B}\,\partial_i\phi)\nonumber\\
 &+\int d^2x \,(\epsilon_{ij} \mathbf{A}\,\partial_j\Pi)(T^{-1}\,\epsilon_{ik} \mathbf{A}\,\partial_k\Pi)
\end{align}
one has
\begin{align}
&\int d^2x (\mathbf{B}\,\partial_i\phi)(T^{-1}\,\mathbf{B}\,\partial_i\phi) = a^2
 \int d^2x \,\phi \Box \phi
\nonumber\\
&\int d^2x \,(\epsilon_{ij} \mathbf{A}\,\partial_j\Pi)(T^{-1}\,\mathbf{B}\,\partial_i\phi) = 0
\nonumber\\
&\int d^2x \,(\epsilon_{ij} \mathbf{A}\,\partial_j\Pi)(T^{-1}\,\epsilon_{ik} \mathbf{A}*\partial_k\Pi) =
\int d^2x \,\Pi^2
\end{align}

To get the third line result we have chosen ${\mathbf{A}= T^{1/2}*\hat T^{-1}}$.

With all this one has for the dual Hamiltonian
\begin{align}
H&=\frac{1}{2} \int d^2x \left( E_i*T^{-1}*E_i + B*T^{-1}*B\right)\nonumber \\
&= \frac{1}{2} \int d^2x \left( \Pi^2 + a^2 \phi \Box \phi  -\frac{a^2}{4 \alpha^2} \phi \nabla^2\phi \right)\nonumber \\
&= \frac{1}{2} \int d^2x \left( \Pi^2 - a^2 \dot \phi^2 + a^2(1-\frac{1}{4 \alpha^2}) \phi \nabla^2\phi \right)
\label{jamil}
\end{align}
so that, choosing the arbitrary parameter $a$ as $a=2\alpha$, we get
\begin{align}
H&=\frac{1}{2} \int d^2x \left( \Pi^2 - 4\alpha^2 \dot \phi^2 + a^2(1-\frac{1}{4 \alpha^2}) \phi \nabla^2\phi \right)\nonumber
\end{align}
Using  $\partial H/\partial \Pi = \dot \phi$ one gets
$\dot \phi =\Pi$ leading to
\be
H=\frac{(1-4\alpha^2)}{2}\int \left(\Pi^2 + \phi(- \nabla^2)\phi\right)
\label{freH}
\ee
which is the Hamiltonian for a massless boson. A field $\phi$ redefinition together with a consistent choice of regularization parameter $\alpha$ leads to the free field massless Lagrangian
 \be
L_\phi =  \int d^3x \,(\partial_\mu\phi)^2
\label{54}
 \ee

With the choice of $\mathbf{A}$, $\mathbf{B}$ y $\mathbf{C}$ that we made, the electric and magnetic fields, as given by eq.\eqref{gauss}, become
\begin{eqnarray}
E_i &=&T^{1/2}\left( \epsilon_{ij}  \hat T^{-1} \partial_j \Pi + 2\alpha T \hat T^{-1}\partial_i\phi\right) \\
B &=&T^{1/2} \hat T\, \phi
\end{eqnarray}
We can also express the generator of spatial translations of the theory  in terms  of $\phi$ finding that it coincides with that of a scalar field,
\be
P^i=\int d^2x \,\epsilon^{ij} E^j\,T^{-1}\,B = \int d^2x\, \Pi \,\partial_i\phi
\ee
Massless representations of the three-dimensional Poincar\'e group describe particles with no spin \cite{Ba}. This  implies that the parity violating massless Dirac theory as well as its dual bosonic theory with Hamiltonian \eqref{freH}  describes spinless fermion and bosons respectively (see ref.\,\cite{DJT} for a discussion on this issue).

\section{Discussion}

We summarize in Figure 1 the duality relations that we established in this work  between generating functionals for current correlation functions of a free massless fermion model in $d=3$ space-time dimensions and several bosonic models. We have not included in the graph the connection found at the end of section 3 with a higher-derivative theory involving purely scalars since in that case the scalar field completely decouples from the external source. Although in that case the scalar can be ignored, the dependence  of the complete bosonic generating function on the source (which can be thought as an electromagnetic potential) leads to the correct massless fermions current algebra.

Concerning the dualities represented in the graph, they extend, for the massless case, those obtained in the massive case. Indeed, for $m\ne 0$ fermions in $d=3$ dimensions the connection with topologically massive gauge theory (i.e, Maxwell-Chern-Simons theory)  was established in \cite{FS}-\cite{LNS}  the mass in the fermionic theory being related to the Chern-Simons  coefficient.  As already shown in \cite{DJ}, Maxwell-Chern-Simons theory is in turn dual to a self-dual theory  and this
closes the triangle in Figure 1 joining the three generating functionals. Interestingly enough, the role of the  mass in the bosonic models dual to massive fermions,  is played in the
massless case by the square-root of the $d=3$ Laplacian, an operator that can be rigorously defined  leading to non-local bosonic theories that are the natural extension of the Maxwell-Chern-Simons and self-dual theories. As mentioned at the end of section 3, canonical quantization of this kind of non-local theories can be formulated preserving causality, as discussed in detail in \cite{Amaral}.

An important point in the dualities discussed in the previous paragraph concerns the fact that although the dual  bosonic actions were calculated within a quadratic approximation, the parity violating part, which is the one leading to the correct current algebra, does not receive higher order corrections so that the bosonization rule \eqref{rule0} is in this sense an exact one.

\begin{figure}[ht]
\centerline{\hspace{4cm} \begin{tikzpicture}[sloped]
  \node (a) at ( 0,0) [bag] {
$ \hspace{-0.4cm}
{\!\!Z_{fer}[s] =\!\int\!\! D\bar{\psi}D\psi \exp\left(-S_{fer}[\bar \psi,\psi;s]\right)}$};
  \node (b) at ( 7,2.2) [bag] {$\hspace{-2cm}{Z_{TMCS}[s]\! = \!\int \!DA_\mu \exp\left(-S_{TMCS}[A;s]\right)}$};
  \node (c) at ( 7,-2.2) [bag] {{\vspace{0.5 cm}$\hspace{-2cm}{Z_{TSF~}[s]\! = \!\int \! Da_\mu \exp\left(-S_{TSD}[a;s]\right)}$}};
  \node (d) at (7,-4.4) [bag]  {$\hspace{-2cm}{Z_{\phi}[s]\! = \!\int \! D\phi \exp\left(-S_{eff}[\phi;s]\right)}$};
  \node (e) at (7,-4.1) [bag] {$\hspace{2cm}$~~~};
  \node (f) at (7,2) [bag] {~~~};
  \draw [<->] (a) to node [above] {} (b);
  \draw [<->] (a) to node [below] {} (c);
 \draw [<->] (b) to node [below] {} (c);
 \draw [<->] (a) to node [below] {} (d);
 \draw[-latex,color=blue]
        (f) .. controls +(right:10.2cm) and
                                +(right:3cm) ..
                                  node[near end,above right,color=black] {}
        (e);
\end{tikzpicture}}
\caption{Dualities between the free fermion model and the bosonic dual models.}
\label{tree}
\end{figure}
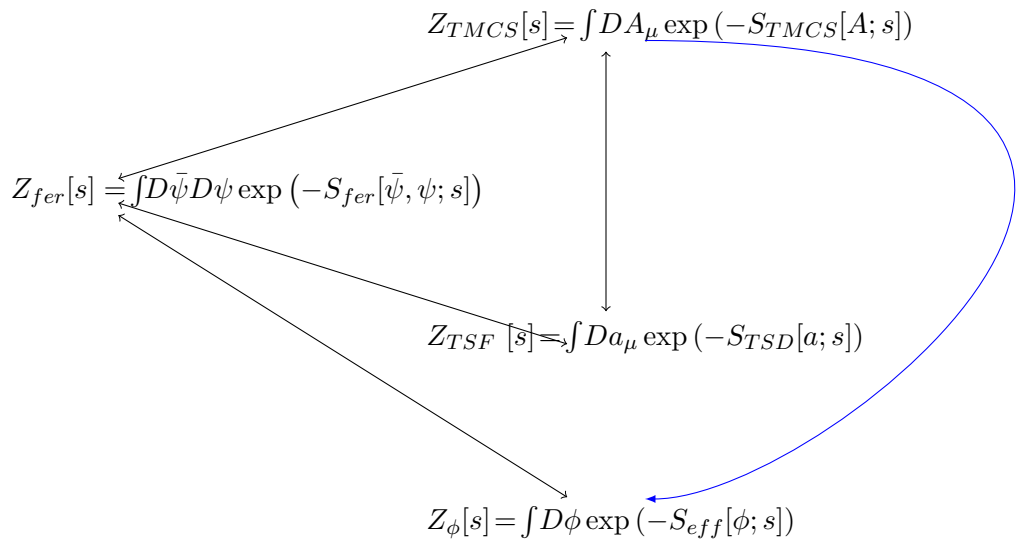

Section 4 was dedicated to find still another duality between $d=3$ fermion and boson models following a procedure similar to that developed in \cite{DJT},\cite{DJ} to prove that topologically massive gauge  theory and Self-dual theory can be written in terms of a canonical, free, massive field $\phi$ with a Lagrangian that looks as that of  a spinless field. In that case a careful analysis of the Lorentz generators as functionals of $\phi$ shows that the spin is in fact $\pm 1$. Interestingly enough,  the square root of the (spatial) Lagrangian plays a central role in this massive case in finding the connection between the vector field $A_\mu$ and the $\phi$ field  \cite{DJT}.

We were able to extend the procedure described above to the case of the massless bosonic theories that we prove to be equivalent to the massless free fermion theory and also in this case   the square root of two-dimensional and three dimensional  Laplacians enter into the game. In this way we were able to write the massless bosonic model with a vector field $A_\mu$ and dynamics governed by a Maxwell-like Chern-Simons action in terms of single field $\phi$ which in this case, due to peculiarities of the Lorentz group in $d=3$ dimensions, has spin $0$, as it is the case for $d=3$ massless fermions.

The duality between the $d=3$ massless fermion Lagrangian and the bosonic one given in eq.\eqref{54} can be seen as the $N=1$ version of the bosonization $1/N$ results obtained in \cite{A} based in the AdS/CFT conjecture. Since the path-integral approach is particularly adequate  to treat $d=3$ non-Abelian bosonization, one could expect that the extension of the results for massive fermions \cite{GMNS} should lead, in the massless case, to results analogous to those presented in \cite{A} concerning the equivalence, for large $N$, of the theory of $N$ massless scalars coupled to a  level $k$  $U(N)$ Chern-Simons term  with with the Legendre transform of a theory of $k$ massless fermions coupled to a level $N$ $U(k)$ Chern-Simons term. Moreover, the authors of \cite{A} conjecture that the fermionic/bosonic duality is valid for finite values of $N$, making our results for $N=1$ consistent with that claim. We hope to discuss this issue in a forthcoming work.

\vspace{.5cm}

\noindent\underline{Acknowledgments}:   This work was supported by  CONICET  , ANPCYT , CIC  and UNLP, Argentina.

\end{document}